\documentclass[11pt,epsf]{article}

\usepackage{epsfig}
\usepackage{amssymb}
\usepackage{graphicx}
\usepackage{color}
\usepackage{subfigure}

\begin{document}

\baselineskip 6mm
\renewcommand{\thefootnote}{\fnsymbol{footnote}}


\newcommand{\nc}{\newcommand}
\newcommand{\rnc}{\renewcommand}



\newcommand{\tcb}{\textcolor{blue}}
\newcommand{\tcr}{\textcolor{red}}
\newcommand{\tcg}{\textcolor{green}}


\def\be{\begin{equation}}
\def\ee{\end{equation}}
\def\ba{\begin{array}}
\def\ea{\end{array}}
\def\bea{\begin{eqnarray}}
\def\eea{\end{eqnarray}}
\def\nn{\nonumber\\}


\def\ct{\cite}
\def\la{\label}
\def\eq#1{(\ref{#1})}


\def\a{\alpha}
\def\b{\beta}
\def\g{\gamma}
\def\G{\Gamma}
\def\d{\delta}
\def\D{\Delta}
\def\ep{\epsilon}
\def\e{\eta}
\def\ph{\phi}
\def\Ph{\Phi}
\def\ps{\psi}
\def\Ps{\Psi}
\def\k{\kappa}
\def\l{\lambda}
\def\L{\Lambda}
\def\m{\mu}
\def\n{\nu}
\def\th{\theta}
\def\Th{\Theta}
\def\r{\rho}
\def\s{\sigma}
\def\S{\Sigma}
\def\ta{\tau}
\def\o{\omega}
\def\O{\Omega}
\def\pr{\prime}


\def\half{\frac{1}{2}}

\def\goto{\rightarrow}

\def\na{\nabla}
\def\grad{\nabla}
\def\curl{\nabla\times}
\def\div{\nabla\cdot}
\def\pa{\partial}

\def\bra{\left\langle}
\def\ket{\right\rangle}
\def\lb{\left[}
\def\lc{\left\{}
\def\ls{\left(}
\def\lp{\left.}
\def\rp{\right.}
\def\rb{\right]}
\def\rc{\right\}}
\def\rs{\right)}
\def\fr{\frac}

\def\vac#1{\mid #1 \rangle}


\def\td#1{\tilde{#1}}
\def\check{ \maltese {\bf Check!}}


\def\Tr{{\rm Tr}\,}
\def\det{{\rm det}}


\def\bc#1{\nnindent {\bf $\bullet$ #1} \\ }
\def\ch {$<Check!>$ }
\def\ss {\vspace{1.5cm}}

\begin{titlepage}

\hfill\parbox{5cm} { }

\vspace{25mm}

\begin{center}
{\Large \bf Holographic meson mass splitting in the Nuclear Matter}

\vskip 1. cm
{Bum-Hoon Lee$^{a,b}$\footnote{e-mail : bhl@sogang.ac.kr},
 Shahin Mamedov$^{c,d}$\footnote{e-mail : sh.mamedov62@gmail.com},
Siyoung Nam$^a$\footnote{e-mail : stringphy@gmail.com} and
  Chanyong Park$^b$\footnote{e-mail : cyong21@sogang.ac.kr}
 }

\vskip 0.5cm

{\it $^a\,$ Department of Physics, Sogang University, Seoul 121-742, Korea}\\
{ \it $^b\,$ Center for Quantum Spacetime (CQUeST), Sogang University, Seoul 121-742, Korea}\\
{\it $^c\,$ Department of Physics, Mimar Sinan Fine Arts University, Bomonti, 34380, Istanbul, Turkey} \\
{\it $^d\,$ Institute for Physical Problems, Baku State University, Z.Khalilov str. 23, Baku, AZ-1148, Azerbaijan} \\

\end{center}

\thispagestyle{empty}

\vskip1cm


\centerline{\bf ABSTRACT} 

\vspace{1cm}
We study the holographic light meson spectra and their mass splitting in the nuclear medium. 
In order
to describe the nuclear matter, we take into account the thermal charged AdS geometry
with two flavor charges, which can be reinterpreted as the number densities of proton
and neutron after some field redefinitions. We show that the meson mass splitting occurs
when there exists the density difference between proton and neutron. 
Depending on the flavor charge,
the mass of the positively (negatively) charged meson increases (decreases) as the
density difference increases, whereas the neutral meson mass is independent of the
density difference. In the regime of the large nucleon density with a relatively large
number difference between proton and neutron, we find that negatively charged pion
becomes massless in the nuclear medium, so the pion condensate can occur.
We also investigate the binding energy of a heavy quarkonium in the nuclear medium,
in which the binding energy of a heavy quarkonium becomes weaker as the density difference
increases.



\end{titlepage}

\renewcommand{\thefootnote}{\arabic{footnote}}
\setcounter{footnote}{0}



\section{Introduction}

After the RHIC and LHC experiments, there were huge amount of efforts to explain
the strongly interacting quark-gluon plasma. A related big issue in the strongly interacting
quantum chromodynamics (QCD) is to understand the nuclear matter and its physical properties
like the
symmetry energy. Since the analytic method is not available in the strong coupling regime
and the numerical method, the so-called lattice QCD, in the dense medium has a sign problem, 
a new method is required to investigate the strongly interacting nuclear
matter. Recently, after the Maldacena's conjecture \cite{Maldacena:1997re,Gubser:1998bc,Witten:1998qj}, many peoples have tried to understand
the strongly interacting QCD-like theory following the spirit of the gauge/gravity duality. 
This new method is widely believed to be a powerful tool in understanding the strongly
interacting system like the QCD and condensed matter theory \cite{Aharony:1999ti,Klebanov:2000me,Horowitz:2006ct,Natsuume:2007qq,Erdmenger:2007cm,Sachdev:2010ch}. 

In order to understand the strongly interacting QCD, various holographic models have been
constructed \cite{Erlich:2005qh,Karch:2006pv,Sakai:2004cn,Sakai:2005yt}. One of them is the hard wall model \cite{Erlich:2005qh}-\cite{Park:2011zp}, in which an IR cut off is introduced 
to explain the quark confinement holographically. Although the Regge behavior is not 
described well, this model gives the good qualitative behavior of the meson spectra comparable to the experiment data. In the hard wall model, the asymptotic AdS space is taken into account
as a background geometry, which in the string theory is dual to the ${\cal N}=4$ 
super Yang-Mills (SYM) theory. The most general asymptotic AdS geometry is, 
if there is no charge hair,
the Schwarzschild AdS (SAdS) black hole whose dual theory
is defined at finite temperature. The pure AdS space, then, corresponds to the zero temperature
limit whose Euclidean version is sometimes called the thermal AdS (tAdS) space. 
After introducing a
hard wall to the pure AdS space, it was shown that there exists a Hawking-Page transition
between the tAdS space and SAdS black hole \cite{Herzog:2006ra}. 
Following the AdS/CFT correspondence, this Hawking-Page transition 
represents the deconfinement phase transition of the dual theory in the zero density limit.

These works have been further generalized to the case of a dense medium \cite{Lee:2009bya,Park:2009nb,Jo:2009xr,Park:2011zp} . 
The density in the dual field theory can be mapped to a bulk gauge field. 
The gravitational backreaction of the gauge field modifies the
background geometry from the SAdS black hole to  Reissner-Nordstrom AdS (RNAdS)
black hole at finite temperature and from the tAdS space to thermal charged AdS (tcAdS) space 
in the zero temperature limit. The tcAdS geometry is the same as the RNAdS black hole
with $m=0$. In general, the tcAdS space has a naked singularity at the center. In the hard
wall model, however, the IR cut off prevents all bulk quantities from approaching 
to the singular point, so the naked singularity is not harmful.  As a result, 
the confining phase of the dense medium at zero temperature can be holographically 
mimicked by the tcAdS space. 
In the RNAdS black hole, there is another
zero temperature limit described by the extremal RNAdS black hole. 
Since it has a black hole horizon, the dual theory is still in the deconfining phase. 
It was shown that there is also a Hawking-Page transition \cite{Lee:2009bya}. In the low density and low
temperature, the tcAdS space corresponding to the confining phase is preferable,
while the RNAdS black hole is dominant in the high temperature and high density regime
similar to the QCD. This charged geometry can be further generalized to the case
with two flavor charges. To do so, the U(1) gauge group has been lifted to the U(2) case, in which
the combinations of two Cartan subgroup elements are mapped to the chemical potentials or 
number density of u- and d-quarks \cite{Park:2011zp}. In the confining phase, since 
the fundamental excitations are not quarks but nucleons, 
the appropriate combinations of such bulk gauge fields were reinterpreted as the density
of nucleons. 
On this background geometry, the symmetry energy caused by the nucleon number difference 
has been studied in which the asymmetry of the nucleon
density increases the free energy of system. 

Recently, it was shown that pions can be condensed in the isospin matter which has only 
the isospin chemical potential without the quark chemical potential \cite{Albrecht:2010eg}. 
In order to describe the nuclear matter, we need to improve the isospin matter. 
In this paper, we will investigate the meson spectra in the nuclear matter
whose dual geometry is given by the tcAdS with an IR cutoff.
In the nuclear matter the meson spectra with a flavor charge are 
usually affected by the nucleon density due to the flavor charges of background nucleons.
We show that, when the ratio of the total number density and the number difference of nucleons
is fixed, the mass of a positively (negatively) charged meson increases (decreases)
with the increasing total nucleon number density. Also, we find that the mass of a heavy 
quarkonium slightly decreases when the total nucleon number density increases.

The rest of the paper is organized as follows: In Sec. 2, we explain the dual background geometry
of the nuclear matter together with our conventions, in which we regard the $U(2)_L \times U(2)_R$ flavor symmetry group for describing proton and neutron. After turning
on the chiral condensate in Sec. 3, we investigate the light meson spectra 
by solving the linearized equations of various bulk fluctuations. In Sec. 4, the binding 
energy of a heavy quarkonium in the nuclear matter is studied. Finally, we finish
our work with some concluding remarks in Sec. 5.

\section{Nuclear medium in the hard wall model}

First, we study the dual geometry of the nuclear matter with introducing our notations.  
We start with the following action 
\be \la{Act:org}
S = \int d^5 x \sqrt{- G} \lb \frac{1}{2 \k^2} \ls  {\cal R} - 2 \L \rs  - \frac{1}{4 g^2} \Tr \ls F^{(L)}_{MN} F^{(L)MN} + F^{(R)}_{MN} F^{(R)MN} \rs  \rb  ,
\ee
where the cosmological constant is given by $\L = - 6/R^2$.
The superscripts, $(L)$ and $(R)$, represent the left and right part of the $U(2)_L \times U(2)
_R$ flavor symmetry group, where 
\bea
F^{(L)}_{MN} &=& \pa_M L_N - \pa_N L_M - i \lb L_M , L_N \rb , \nn
F^{(R)}_{MN} &=& \pa_M R_N - \pa_N R_M - i \lb R_M , R_N \rb. 
\eea
In order to consider the nuclear matter holographically, we turn only on the time-component 
gauge fields of the diagonal elements because non-zero values of them give rise to the definite 
meanings in the dual gauge theory. 
Denoting the non-zero gauge field as $L^{0}_0$, $L^{3}_0$, $R^{0}_0$, and $R^{3}_0$, 
where the superscript and subscript imply a gauge group index and time 
component respectively,  the $U(2)_L \times U(2)_R$ flavor symmetry group is broken to 
$U(1)_L^2 \times U(1)_R^2$. Following the AdS/CFT correspondence, the boundary values of 
these gauge fields correspond to the chemical potentials, which are the sources of 
the density operators.
Before finding the background geometry, we first summarize our notations for 
later convenience
\bea
M, N &=& 0, \cdots, 4 \quad \ls {\rm bulk \ directions } \rs , \nn
\m, \n &=& 0, \cdots, 3 \quad \ls {\rm  boundary  \ directions } \rs , 
\nn 
m, n &=& 1, 2, 3 \quad \ls {\rm spatial \ directions \ of \ the \ boundary \ space } \rs , \nn
a, b &=& 0, \cdots, 3 \quad \ls {\rm  indices \ of } \  U(2) \ {\rm  flavor \ group } \rs , \nn
i,j &=& 1, 2, 3 \quad \ls  {\rm  indices \ of } \  SU(2) \subset  U(2) \rs .
\eea

In order to know the background geometry, we need to decompose the above reduced Abelian gauge fields into the symmetric and anti-symmetric part
\bea		\la{vectorbaryon}
V^a_0 &=& \frac{1}{2} \ls L^{a}_0 + R^{a}_0 \rs , \nn
A^a_0 &=& \frac{1}{{2}} \ls L^{a}_0 - R^{a}_0 \rs  .
\eea
Here, we use the following normalizations 
\be
L_M \equiv L^{a}_M \  T^{a} \quad {\rm and} \quad R_M \equiv R^{a}_M \  T^{a} , \nn
\ee
with 
\be
T^{a} = \lc \frac{{\bf 1}}{{2}}, \frac{\s^i}{{2}} \rc  ,
\ee
where ${\bf 1}$ and $\s^i$ are the identity and Pauli matrices. 
In \eq{vectorbaryon}, the boundary values of $V^{0}_0$ and $V^{3}_0$ map to the 
chemical potentials of quark and isospin, respectively. Although we can also introduce the chemical 
potential for $A^{a}_0$, we concentrate on the case invariant under exchanging of 
the left and right gauge groups. In other words, $L^{a}_0 = R^{a}_0$.  
Then, the action can be rewritten in terms of new variables as
\be		\la{backaction}
S = \int d^5 x \sqrt{- G} \lb \frac{1}{2 \k^2} \ls  {\cal R} - 2 \L \rs  - \frac{1}{4 g^2} \ls F^{0}_{MN} F^{0 MN} + F^{3}_{MN} F^{3 MN} \rs  \rb ,
\ee
where $F^{0}_{MN}$ and $F^{3}_{MN}$ are the field strengths of two diagonal Abelian gauge fields
\be
F^a_{MN} = \pa_M V^a_N - \pa_N V^a_M  ,
\ee
where $a$ is $0$ or $3$.
This action was used in studying the holographic symmetry energy \cite{Park:2011zp}.

For more concrete interpretation of the bulk gauge fields in the dual theory, we 
introduce the u- and d-quark gauge fields as
\bea
A^{u}_0 &=& \frac{1}{\sqrt{2}} \ls V^{0}_0 + V^{3}_0 \rs , \nn
A^{d}_0 &=& \frac{1}{\sqrt{2}} \ls V^0_0 - V^3_0 \rs .
\eea
Then, the u- and d-quark gauge fields 
describe holographically the quark matters composed of $u$- and $d$-quarks.  
It is worth noting that the isospin matters with two flavors can be
similarly described by taking $V^0_0 = 0$ and $V^3_0$ as a constant \cite{Son:2000xc}.
The action \eq{backaction} then can be reexpressed as
\be		\la{act:quark}
S = \int d^5 x \sqrt{- G} \lb \frac{1}{2 \k^2} \ls  {\cal R} - 2 \L \rs  - \frac{1}{4 g^2} \ls F^{u}
_{MN} F^{u MN} + F^{d}_{MN} F^{d MN} \rs  \rb ,
\ee
where slightly different conventions from \cite{Park:2011zp} are used. 
The relations between the above gauge fields can be summarized as
\bea 		\la{back:gauge field}
&& L^{0}_0 = R^{0}_0 =  V^0_0 = \fr{1}{\sqrt{2}} \ls A^u_{0} + A^d_{0} \rs , \nn
&& L^{3}_0 = R^{3}_0 =  V^3_0 = \fr{1}{\sqrt{2}} \ls A^u_0 - A^d_0 \rs .
\eea
If we assume that all variables depend only on the radial coordinate $z$, 
the thermal charged AdS (tcAdS) geometry, which is a non-black brane solution satisfying the Einstein and  Maxwell equations simultaneously,
is given by
\be		\la{backgeometry}
ds^2 =  \frac{R^2}{z^2}  \ls - f(z) dt^2 + \frac{1}{f(z)} dz^2  + d \vec{x}^2
\rs ,
\ee
with
\be
f(z) = 1 + \sum_{\a=u,d} q_{\a}^2 z^6 ,
\ee
where $R$ is the AdS radius. 
The bulk gauge field satisfying the Maxwell equations becomes
\be			\la{sol:quarkgauge}
A^{\a}_0 = 2 \pi^2 \m_{\a} - Q_{\a} z^2 ,
\ee
where $\m_{\a}$ and $Q_{\a}$ corresponds to the chemical potential and number
density of u- and d-quark respectively. 
Note that $q_{\a}$ in the tcAdS space is related to the number density
$Q_{\a}$ 
\be 	\la{rel:para}
q_{\a}^2  = \frac{2 \k^2}{3 g^2 R^2}  Q_{\a}^2  .
\ee

There are several remarks associated with the background geometry. 
According to the AdS/CFT correspondence, the tcAdS geometry
can be interpreted as a dual geometry describing the zero temperature gauge theory
without the quark confinement. In order to study the meson spectra in the
confining phase, we need a proper prescription for the quark confinement. 
In the bottom-up model, there are two different ways for explaining the confining behavior. One 
is the hard wall model \cite{Erlich:2005qh,Herzog:2006ra} and the other is the soft wall model 
\cite{Karch:2006pv,Colangelo:2010pe}. 
Here, we concentrate on the hard wall model. In the hard wall model, 
an IR cut-off or hard wall is introduced at a finite radius, $z=z_{IR}$, to
explain the confining behavior. 
Although the background geometry \eq{backgeometry} has a curvature singularity 
at $z=\infty$, the hard 
wall prevents all bulk quantities from approaching to the singularity. Consequently, the 
singularity at $z=\infty$ is not harmful at least in the hard wall model. 
When regarding a deconfining phase, a well-known dual geometric solution is
given by the Reinssner-Nordstron AdS (RNAdS) black brane, which has also its own zero 
temperature limit called the extremal black hole. 
Due to the existence of two different zero temperature geometries, we should think of which 
the more preferable geometry is. In \cite{Lee:2009bya,Park:2011zp}, it has been shown 
that the non-black hole geometry is more preferable in the low temperature and low density 
regime, and that there exists a deconfinement phase transition even at zero temperature. 
Above the critical density, the 
extremal RNAdS black hole describing the deconfining phase is dominant. Since we are 
interested in the meson spectra in the confining phase, the tcAdS
geometry in \eq{backgeometry} would be a good background for such an investigation.

Following the AdS/CFT correspondence, the on-shell gravity action with an Euclidean signature, after an appropriate holographic renormalization, corresponds to the thermodynamic energy of the dual gauge theory. 
If we impose the Dirichlet boundary condition on the bulk gauge field, which determines the Euclidean action in terms of the chemical potentials, the thermodynamic energy reduces to the grand potential defined in the grand canonical ensemble 
\be
\O (\m_u,\m_d) = - \frac{R^3 V_3}{ \k^2} \ls \frac{1}{z_{IR}^4} + \frac{2 \k^2}{3 g^2 R^2} \
(Q_u^2 + Q_d^2) \ z_{IR}^2 \rs ,
\ee
where $Q_u$ and $Q_d$ are functions of $\m_u$ and $\m_d$. At this stage,
the number density is not determined as a function of the chemical potential. 
In order to find the relation between $Q_{\a}$ and $\m_{\a}$, we should also take into account
the thermodynamic relation between the grand potential and free energy. If we impose
the Neumann boundary condition instead of the Dirichlet boundary condition, the corresponding
thermodynamic energy is given by the free energy of the canonical ensemble. 
In this case, we should add a boundary term to the original action \eq{act:quark} 
for fixing the number density. Furthermore, since the chemical potential and number density are 
conjugate  variables, these two thermodynamic energies must be related to each other by the 
Legendre transformation \cite{Lee:2009bya,Park:2009nb,Jo:2009xr,Park:2011zp}.  The requirement of the well-defined Legendre transformation 
determines the number density as a function of the chemical potential or vice versa
\be        \la{res:relchden}
Q_{\a} = \frac{3 \pi^2}{z_{IR}^2}  \ \m_{\a} .
\ee 
where $1/g^2 = N_c / (4 \pi^2 R) $ and $1/(2 \k^2) = N_c^2 / (8 \pi^2 R^3)$ are used
and $N_c$ represents the rank of the gauge group of the dual theory. 
Using this result, as one expected,
the total number of $u$- or $d$-quarks $N_{\a}$ at a given volume $V_3$
becomes
\be
N_{\a} \equiv - \frac{\pa \O}{\pa \m_{\a}} = V_3 \  N_c \ Q_{\a} .
\ee

In the confining phase, the fundamental excitations are not quarks but nucleons due to the confinement. Therefore, it is required to translate the
quark number density \eq{back:gauge field} into the nucleon number density. 
Since the numbers of $u$- and 
$d$-quarks are conserved, they can be easily represented in terms of 
the nucleon number densities 
\be
Q_u =  2 Q_P + Q_N     \quad  {\rm and} \quad Q_d = Q_P + 2 Q_N ,
\ee  
where $Q_P$ or $Q_N$ denotes the number density of proton or neutron respectively. 
Then, $f(z)$,  $V_0^0 (z)$ and $V_0^3 (z)$ can be rewritten as
\bea
f (z) &=& 1 +\frac{3 Q^2 \k^2 }{g^2 R^2} z^6 + \frac{D^2 \k^2}{3 g^2 R^2} z^6 , \nn
V_0^0 (z) &=&  \frac{Q}{\sqrt{2}}  \ls 2 z_{IR}^2 -  3 z^2  \rs  , \nn
V_0^3 (z) &=&  \frac{D}{3\sqrt{2}}  \ls  2 z_{IR}^2 - 3  z^2  \rs  ,
\eea
where the total nucleon number density and the density difference between proton and neutron
are denoted by $Q \equiv Q_P + Q_N$ and $D \equiv Q_P - Q_N$. 
From now on, we set $D = \a Q$ where $\a$ runs from $-1$ to $1$. In this notation, 
$\a = 0$ implies that the nuclear matter has the same number of proton and neutron,
while $\a= \pm 1$ means that the nuclear matter is composed of only one species of matter, 
proton or neutron.

Finally, the isospin matter with two flavors can be taken into account
by setting $Q_{\a} = 0$ in \eq{sol:quarkgauge} \cite{Son:2000xc,Parnachev:2007bc,Kim:2007gq,Aharony:2007uu}. In this case the tcAdS space reduces to the tAdS space, which is the proper background geometry for the isospin matter.
Moreover, since the isospin matter by definition is described only by the isospin chemical
potential without the quark (or nucleon) chemical potential, $V^0_0$ should be zero.
Then, $V^3_0$ is fixed to the difference of the quark isospin chemical potentials
$V^3_0=\sqrt{2} \pi^2 (\m_{u} - \m_{d})$.
In the confining phase, $V^3_0$ again can be reexpressed in terms of the isospin chemical potentials of nucleons, $\m_P$ and $\m_N$, where $P$ and $N$ imply proton and neutron respectively.
If we set naively the isospin chemical potential of nucleons to be the sum of
quark chemical potetials, $\m_P = 2 \m_u + \m_d$ and $\m_N = \m_u + 2 \m_d$, we can
find from \eq{back:gauge field} 
\be			\la{res:isospinche}
V^3_{iso}= \sqrt{2} \ \pi^2  (\m_P - \m_N) ,
\ee
which corresponds to the isospin chemical potential in the isospin matter. 
 
\section{Light meson spectra in the nuclear medium}

Now, we turn on various bulk fluctuations representing mesons of the dual gauge theory
\be \la{Act:fluc}
S_f = - \int d^5 x \sqrt{- G}  \  \Tr \lb  \left| D \Ph \right|^2 + m^2 \left| \Ph \right|^2 + \frac{1}{4 g^2} \ls F^{(L)}_{MN} F^{(L)MN} + F^{(R)}_{MN} F^{(R)MN} \rs \rb  ,
\ee
where $D_M \Ph = \pa_M \Ph - i L_M \Ph + i \Ph R_M$. The fluctuations of the gauge fields
and complex scalar fields can be described by
\bea   \la{ans:mesonfluctuations}
&& L^a_M = \bar{L}^a_M + l^a_M \quad {\rm and} \quad R^a_M = \bar{R}^a_M + r^a_M , \nn
&& \Ph = {\cal N} \ph \ {\bf 1} \ e^{i \sqrt{2} \pi^a T^a} ,
\eea
where ${\cal N}$ is a normalization constant and for simplicity we set ${\cal N}=1$. 
In the above equation, $\bar{L}_M$ and $\bar{R}_M$ are the background gauge fields appearing in \eq{back:gauge field}, whereas
$l^i_M$, $r^i_M$ and $\pi^i$ are fluctuations corresponding vector, axial vector and pseudoscalar mesons, respectively. Since we are interested in the meson spectra of the $SU(2)$ flavor
group, we only turn on the fluctuations of the $SU(2)$ sector 
($l^i_{\m}, \ r^i_{\m}, \ \pi^i \ne 0$)
and set others to be zero ($l^0_{\m} = r^0_{\m} = \pi^0 = 0$). We further consider all non-zero fluctuations
as functions of $t$ and $z$ only, which is always possible by taking the rest frame.

Related to the modulus of the complex scalar field dual to the chiral condensate, 
we regard $\ph$ as a background field without considering the gravitational backreaction of it. 
The gravitational backreaction effect of $\ph$ in the nuclear matter without the isospin effect 
has been studied numerically in \cite{Lee:2010dh}. The equations of motion for $\ph$ becomes
\be
0 = \frac{1}{\sqrt{-G}} \pa_z \ls \sqrt{-G} g^{zz} \pa_z \ph \rs + 3 \ph ,
\ee
where $m^2 = -3/R^2$ and we set $R=1$. The solution of the modulus is given by
\be
\ph (z) = m_q \ z \ _2 F_1 \ls \frac{1}{6} , \half , \frac{2}{3}, - \frac{\ls D^2 + 9 Q^2 \rs   
z^6 }{3 \ N_c} \rs
 + \s \ z^3 \ _2 F_1 \ls \half, \frac{5}{6},\frac{4}{3},  - \frac{\ls D^2 + 9 Q^2 \rs   
z^6 }{3 \ N_c}\rs ,
\ee 
where two integration constants, $m_q$ and $\s$, correspond to a current quark mass 
and chiral condensate of the dual theory respectively. 
Although $u$- and $d$-quark in nature have slightly different masses, 
here we ignore such a mass difference.
The asymptotic form of this exact solution near the boundary ($z \sim 0$) reduces to
\be
\ph (z) \approx m_q z + \s z^3 + \cdots ,
\ee
where ellipsis implies the higher order corrections.

Now, we take the axial gauge, $l^i_z=r^i_z = 0$ and introduce new variables
\be
v^i_{\m} = \frac{1}{\sqrt{2}} \ls l^i_{\m}  + r^i_{\m}  \rs \quad {\rm and}
\quad a^i_{\m} = \frac{1}{\sqrt{2}} \ls l^i_{\m}   - r^i_{\m}  \rs \quad (i=1,2,3) ,
\ee
where $\m$ implies the space-time coordinates of the boundary space ($\m=0,1,2,3$). These 
new variables are associated with the vector and axial vector meson in the dual gauge theory.
In order to
investigate the meson spectra, we expand the action \eq{Act:fluc} up to quadratic order 
and then solve the linearized equations of motion.  
The action of the fluctuations \eq{Act:fluc} is expanded to
\be
S_{fluc} = S_{v} + S_{as} ,
\ee
where $S_v$ describes the part of the vector fluctuations
\bea		\la{eq:vector}
S_v &=& -  \frac{1}{4 g^2} \int d^5 x \sqrt{- G} \ \lb 
\sum_{i=1}^{3}  \lc  G^{zz} G^{\m \n} \ \pa_z v^i_{\m} \ \pa_z v^i_{\n} + G^{\m \n} G^{m n} \ \pa_{\m} v^i_{m} \ \pa_{\n} v^{i}_{n}  \rc  \rp \nn
&&\lp + \ G^{00} G^{m n} \ \ls V_0^3 \rs^2 \ \ls  v^1_{m} v^1_{n} +  v^2_{m} v^2_{n}  \rs 
+ 2  \ G^{00} G^{m n} \ V_0^3 \  \ls v^1_m \ \pa_0  v^2_{n}   - v^2_{m} \ \pa_0 v^1_n  \rs  \rb ,
\eea
and $S_{as}$ is for the axial vector and pseudoscalar fluctuations 
\bea		\la{eq:axial}
S_{as} &=& -  \frac{1}{4 g^2} \int d^5 x \sqrt{- G} \ \lb  \
 \sum_{i=1}^{3} \ \lc  G^{zz} \ G^{\m \n} \ \pa_z a^i_{\m} \ \pa_z a^i_{\n} 
 + G^{\m \n} \ G^{m n} \ \pa_{\m} a^i_{m} \ \pa_{\n} a^{i}_{n}  \rc \rp \nn
&&   +  \ G^{00}  G^{m n}  \ \ls V_0^3 \rs^2 \ls  a^1_{m} a^1_{n} +  a^2_{m} a^2_{n}  \rs 
+ 2  \ G^{00}  G^{m n} \ V_0^3 \ \ls a^1_m  \ \pa_0  a^2_{n}  -  a^2_m \ \pa_0 a^1_n  \rs  \nn
&& +  \ 4 g^2  \ph^2  \ \sum_{i=1}^3 \lc  G^{\m \n} \ a^i_{\mu} a^i_{\n}  + 
G^{z z} \ \pa_z \pi^i \ \pa_z \pi^i +  G^{\m \n}  \pa_{\m} \pi^i  \ \pa_{\n} \pi^i 
 - 2  \ G^{\m \n} a^i_{\m} \ \pa_{\n} \pi^i  \rc \nn
&& + \ 4 g^2  \ph^2  \ \lc  \ 2 \ G^{00} \ V^3_0 \ \ls \pi^1  \pa_0 \pi^2 - \pi^2  \pa_0 \pi^1 \rs  + 2 \ G^{00} \ V^3_0 \ \ls a^1_0 \ \pi^2 - a^2_0 \ \pi^1 \rs \rp \nn
&&  \lp \qquad  \qquad  \fr{}{}   \lp + \ G^{00} \ \ls V^3_0 \rs^2 \ \ls \pi^1 \pi^1 + \pi^2 
\pi^2  \rs \ \rc \ \rb .
\eea
Here, we use $ V_0^3 = \bar{L}^{3}_0 = \bar{R}^{3}_0$. 
Notice that the 
effect of the background field $V^0_0$ does not appear because the generator $T^0$ always  
commutes with others.

\subsection{Vector mesons}

As shown in \eq{eq:vector}, the vector fluctuations do not mix with the axial 
vector and pseudoscalar fluctuations at least at quadratic order. So we can investigate the 
vector meson spectra from \eq{eq:vector} independently. 
In the bulk, Lorentz symmetry between $t$ and $\vec{x}$ is broken due to the metric factor of 
the background geometry, so the time component of the vector fluctuations behaves differently from the
spatial components. For example, the time component of the vector 
fluctuations describes the change of the chemical potential or density
instead of the meson spectra. As a result, for describing the meson spectra it is
reasonable to consider only the spatial components of fluctuations. In other words, we set the time component fluctuation to be zero, $v^i_0 = 0$.

In order to express $\r$-mesons which are the lowest modes of the vector meson in QCD,
we rewrite the vector fluctuations, in the axial gauge $v_z = 0$, as the following form
\bea
\r^0_{m} &=& v^3_{m}  , \nn
\r^{\pm}_{m}  &=& \frac{1}{\sqrt{2}} \ls v^1_{m}  \pm i v^2_{m}  \rs,
\eea
where $\r^0_{m} $ and $\r^{\pm}_{m} $ are neutral and charged $\r$-meson 
respectively and $\r^-_{m}$
is the complex conjugation of $\r^+_{m}$. 
Taking the Fourier mode expansion at the rest frame
\bea
\r^0_{m} (t,z) = \int \frac{d \o_0}{2 \pi} \ e^{- i \o_0 t} \ \r^0_{m} (\o_0,z) , \nn
\r^{\pm}_{m} (t,z) = \int \frac{d \o_{\pm}}{2 \pi} \ e^{- i \o_{\pm} t} \ \r^{\pm}_{m}
 (\o_{\pm},z) ,
\eea
the equations of motion for $\r$-mesons reduce to
\bea 		\la{eq:vecstpectrum}
0 &=& \pa_z \ls \sqrt{-G} \ G^{zz} G^{m n} \pa_z \r^0_{n} \rs - \o^2_0 \sqrt{-G} \ 
G^{00} G^{m n} 
  \r^0_{n} , \nn
0 &=& \pa_z \ls \sqrt{-G} \ G^{zz} G^{m n} \pa_z \r^{\pm}_{n} \rs - 
\ls \o^2_{\pm} + \ls V_0^3 \rs^2 \mp 2 \o_{\pm} V_0^3 \rs \sqrt{-G} \ G^{00} G^{m n}  
\r^{\pm}_{n} ,
\eea
where $\o_{\pm}$ and $\o_0$ can be identified with masses of $\r^{\pm}_m$ and $\r^0_m$
respectively. 
Although there is no explicit term for the number densities of nucleons, 
they can affect on the meson 
spectrum via a contributions from the background metric.
Substituting the explicit metric components into the above equation, we can find the second
order differential equations governing the vector fluctuations
\bea			\la{eq:vectormeson}
0 &=&   \pa_z^2 \r^0_n 
+ \frac{ -3 N_c + 5 (D^2 + 9 Q^2) z^6 }{  z \lb 3 N_c + (D^2 + 9 Q^2) z^6 \rb} \pa_z \r^0_n 
+ \frac{9 N_c^2 \o_0^2 }{ \lb 3 N_c + (D^2 + 9 Q^2) z^6) \rb^2 }  \r^0_n , \nn
0 &=&   \pa_z^2 \r^{\pm}_n 
+ \frac{ -3 N_c + 5 (D^2 + 9 Q^2) z^6 }{ z \lb 3 N_c + (D^2 + 9 Q^2) z^6 \rb } \pa_z 
\r^{\pm}_n 
+ \frac{9 N_c^2 \ls \o_{\pm} \mp V_0^3 \rs^2 }{ \lb 3 N_c + (D^2 + 9 Q^2) z^6) \rb^2 }  
\r^{\pm}_n  .
\eea

Before studying the meson spectra in the nuclear medium, we consider the isospin matter 
by setting $Q=D=0$ and 
$V_0^3$ to be a constant. 
The equations for the vector meson in the isospin matter leads to
\bea
0 &=&  \pa_z^2  \r^0_n  - \frac{ 1 }{ z} \ \pa_z \r^0_n  + \o_0^2  \  \r^0_n , \nn
0 &=&   \pa_z^2  \r^{\pm}_n - \frac{ 1 }{  z} \ \pa_z^2  \r^{\pm}_n +  \ls \o_{\pm} \mp \sqrt{2} \ \pi^2  (\m_P - \m_N) \rs^2 \ \r^{\pm}_n  .
\eea
Comparing the above two equations,
the masses of the charged $\r$-mesons are related to the neutral one 
\bea			\la{eq:vectormes}
m_{\r^0} &=& \o_0 , \nn
m_{\r^{\pm}} &=& m_{\r^0} \pm  \sqrt{2} \ \pi^2   (\m_P - \m_N) ,
\eea    
which shows the mass splitting of the vector mesons in the isospin matter 
\cite{Parnachev:2007bc,Kim:2007gq,Albrecht:2010eg}. If the chemical potential of proton is larger than  that of neutron,  
the mass of the positively (or negatively) charged $\rho$-meson increases (or decreases). 
In the opposite situation, the mass of $\r^+$ ( or $\r^-$) reversely decreases (or increases).
In addition, the mass splitting of the charged $\r$-mesons increases with the increasing  
chemical potential difference between proton and neutron.
 
In the nuclear matter, we should turn on the non-zero values of $Q$ and $D$. 
The absence of the nuclear matter $Q=0$ represents the zero density limit studied
in \cite{Erlich:2005qh}, in which the background geometry reduces to the tAdS space
and $D$ also vanishes due to the relation $D \le Q$.
From \eq{res:relchden}, we therefore see that the zero density limit of the nuclear matter is different from the isospin matter.
In \eq{eq:vectormeson}, since $V_0^3$ is a function of $z$, we can not directly read the
mass relation unlike the isospin matter case. In order to know the $\r$-meson masses,
we need to solve the above second order differential equations with appropriate
two boundary conditions. After imposing the Dirichlet boundary condition at the
asymptotic boundary and the Neumann boundary condition  at the IR cutoff,
we solve the differential equations numerically. The results is plotted in Fig. 1, which shows 
that the mass splitting becomes large in the dense nuclear matter. Moreover, there exists
a critical density at which $\r^-$-meson becomes massless. Can $\r^-$-meson have
the zero mass in the nuclear medium? To answer this question, we should check
the deconfinement phase transition to see whether this critical density is still in
the confining phase. Following \cite{Lee:2009bya,Park:2009nb,Park:2011zp},
the renormalized free energy of the dual theory in the confining phase is given by
\be 
F_{con} = - \fr{R^3 V_3}{\k^2} \ls \fr{1}{z_{IR}^4} - \fr{2 \k^2}{3 g^2 R^2}  \ls Q_u^2
+ Q_d^2 \rs z_{IR}^2\rs ,
\ee
and in the deconfining phase described by the RNAdS black hole it reduces to
\be
F_{dec} = \fr{R^3 V_3}{\k^2} \ls - \fr{1}{2 z_{h}^4} + \fr{5 \k^2}{3 g^2 R^2}  \ls Q_u^2
+ Q_d^2 \rs z_{h}^2 \rs ,
\ee
where $z_h$ is the black hole horizon which is a function of the Hawking temperature
and charges. At zero temperature, the black hole horizon is related to charges
\be
z_h^6 = \fr{3 g^2 R^2}{ \ls Q_u^2 + Q_d^2 \rs \k^2} 
= \fr{6 g^2 R^2}{ \ls 9 Q^2 + D^2 \rs \k^2} .
\ee
Using this result, the deconfinement phase transition happens at zero temperature 
when the difference of the above two free energies becomes zero
\be
0 = \fr{9}{2 z_h^4} + \fr{1}{z_{IR}^4} - \fr{2 z_{IR}^2}{z_h^6} .
\ee
For $\a = 1/2$, the deconfinement phase transition occurs at $Q=0.1679$,
which implies that before the $\r^-$-meson mass becomes zero the confining phase is
changed into the deconfining phase.

\begin{figure}
\vspace{-1cm}
\centerline{\epsfig{file=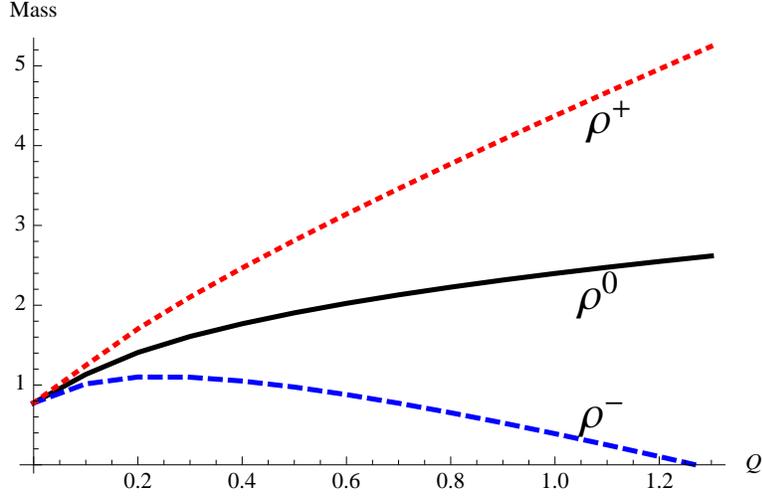,width=10cm}}
\vspace{0cm}
\caption{\small The $\r$-meson mass splitting for $\a = 1/2$, where we take $N_c =3$, 
$R=1$ and $z_{IR}= 1/0.3227$.}
\label{density}
\end{figure}

\subsection{Axial vector mesons}

Let us consider axial vector mesons. In \eq{eq:axial}, there are
several terms including the coupling of the axial gauge field and pseudoscalar fluctuations. 
To get rid of such mixing terms, 
we regard the following gauge transformation 
\be		\la{ans:longitud}
a^i_{\m} = \bar{a}^i_{\m} + \pa_{\m} \chi^i ,
\ee
where $\bar{a}^i_{\m}$ is a transverse fluctuation satisfying 
\be		\la{cond:gauge}
0 = \pa^{\m} \bar{a}^i_{\m} ,
\ee
and the other $\pa_\mu \chi^i$ describes a longitudinal fluctuation. 
We concentrate only on the spatial components of the axial vector
fluctuation $\bar{a}^i_m$ by choosing $\bar{a}^i_0 = 0$, which is due to the same
reason explained in the vector meson case.
Then, the action \eq{eq:axial} in the axial gauge $\bar{a}_z = 0$ can be divided into two parts. One is for the transverse axial
gauge field $\bar{a}^i_m$
\bea
S_{a} &=& -  \frac{1}{4 g^2} \int d^5 x \sqrt{- G} \ \lb 
\sum_{i=1}^{3}  \lc  G^{zz} G^{m n} \ \pa_z \bar{a}^i_{m} \ \pa_z \bar{a}^i_{n} + G^{\m \n} G^{m n} \ \pa_{\m} \bar{a}^i_{m} \ \pa_{\n} \bar{a}^{i}_{n}  \rc  \rp \nn
&&\lp + G^{00} G^{m n} \ls V_0^3 \rs^2 \ls  \bar{a}^1_{m} \bar{a}^1_{n}
 +  \bar{a}^2_{m} \bar{a}^2_{n}  \rs 
 + 2  G^{00} G^{m n} V_0^3  \ls  \bar{a}^1_m \ \pa_0  \bar{a}^2_{n}  - \bar{a}^2_{m}  \
 \pa_0 \bar{a}^1_n   \rs  \rp \nn
&& \lp +  \ 4 g^2  \ph^2  \ \sum_{i=1}^3  G^{m n} \ \bar{a}^i_{m} \bar{a}^i_{n} 
\rb  ,
\eea
where the last term comes from the action of the scalar fluctuations. The other 
has the mixing between the longitudinal axial vector and pseudoscalar fields (see the 
next section). 

For more physical interpretation, we introduce new 
variables $a_{1 m}^0$ and $a^{\pm}_{1 m}$, which denote the lowest axial vector meson,
the so-called $a_1$-meson, 
\bea
a^0_{1 m} &=& \bar{a}^3_{ m}  , \nn
a^{\pm}_{1 m}  &=& \frac{1}{\sqrt{2}} \ls \bar{a}^1_{m}  \pm i \bar{a}^2_{m}  \rs .
\eea
The axial vector mesons are governed by the similar equations to 
$\r$-mesons only except the new mass term caused by the vacuum expectation 
value of the scalar field
\bea
0 &=& \pa_z \ls \sqrt{-G} \ G^{zz} G^{m n} \pa_z a^0_{1 n} \rs - 
\ls G^{00}  \o^2_0 + 4 g^2 \ph^2   \rs  \sqrt{-G} \ G^{m n} 
a^0_{1 n} , \nn
0 &=& \pa_z \ls \sqrt{-G} \ G^{zz} G^{m n} \pa_z a^{\pm}_{1 n} \rs - 
\lb G^{00} \ls \o^2_{\pm} + \ls V_0^3 \rs^2 \mp 2 \o_{\pm} V_0^3 \rs 
+ 4 g^2  \ph^2 \rb \sqrt{-G} \ G^{m n}  a^{\pm}_{1 n} ,
\eea
where we also used the Fourier mode expansion
\bea
a^0_{1 m} (t,z) = \int \frac{d \o_0}{2 \pi} \ e^{- i \o_0 t} \ a^0_{1m} (\o_0,z) , \nn
a^{\pm}_{1m} (t,z) = \int \frac{d \o_{\pm}}{2 \pi} \ e^{- i \o_{\pm} t} \ a^{\pm}_{1m} (\o_{\pm},z) ,
\eea
and $\o_0$ and $\o_{\pm}$ mean the masses of the neutral and charged $a_1$-mesons 
respectively. More explicitly, the differential equations governing the axial vector mesons in the nuclear 
medium reduce to
\bea
0 &=&\pa_z^2 a^0_{1m}
+ \frac{ -3 N_c + 5 (D^2 + 9 Q^2) z^6 }{  z \lb 3 N_c + (D^2 + 9 Q^2) z^6) \rb} \ \pa_z a^0_{1m} \nn
&& + \frac{9 N_c^2 \o_0^2 z^2
-  12 N_c \ g^2  \lb 3 N_c + (D^2 + 9 Q^2) z^6 \rb \ph^2 }{z^2 \lb 3 N_c + (D^2 + 9 Q^2) z^6) \rb^2 }  \ a^0_{1m} , \nn
0 &=& \pa_z^2 a^{\pm}_{1m}
+ \frac{ -3 N_c + 5 (D^2 + 9 Q^2) z^6 }{   z \lb 3 N_c + (D^2 + 9 Q^2) z^6) \rb} \ \pa_z a^{\pm}_{1m} \nn
&& + \frac{9 N_c^2 (\o_{\pm}  \mp V_0^3 )^2 z^2
-  12 N_c \ g^2 \lb 3 N_c + (D^2 + 9 Q^2) z^6 \rb \ph^2 }{z^2 \lb 3 N_c + (D^2 + 9 Q^2) z^6) \rb^2 }  \ a^{\pm}_{1m}  .
\eea
In Fig. 2, after imposing the same boundary conditions used in the previous section, we plot the 
$a_1$-meson masses with $m_q =2.383 MeV$ and $\s = (304 MeV)^3$ used in \cite{Lee:2010dh}.
The mass splitting of $a_1$-meson becomes large in the dense nuclear medium 
like the $\r$-meson case. Similar to $\r^-$-meson, although $a_1^-$-meson becomes
massless at a critical density,
the deconfinement phase transition occurs before arriving at the critical point.

In the isospin matter satisfying $D=Q=0$ and \eq{res:isospinche}, $a_1$-mesons have the following mass specta
\bea			\la{eq:axialvectormes}
m_{a_1^0} &=& \o_0 , \nn
m_{a_1^{\pm}} &=& m_{a_1^0} \pm  \sqrt{2} \ \pi^2   (\m_P - \m_N) .
\eea 
The mass difference between a neutral and charged $a_1$-meson shows the same result obtained in the $\r$-meson spectra.

\begin{figure}
\vspace{-1cm}
\centerline{\epsfig{file=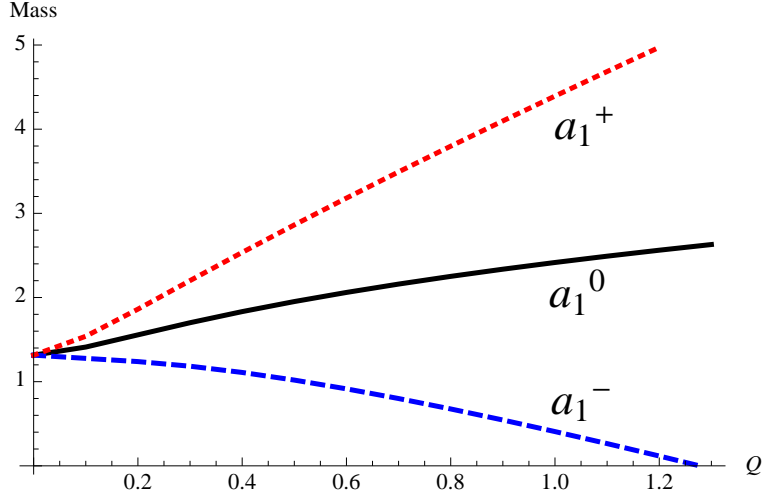,width=10cm}}
\vspace{0cm}
\caption{\small The $a_1$-meson mass splitting for $\a = 1/2$, where we take 
$m_q =2.383 MeV$, $\s = (304 MeV)^3$, $N_c =3$ and $R=1$.}
\label{density}
\end{figure}

\subsection{Pseudoscalar mesons}

The mass spectra of the pseudoscalar mesons are described by the action including the coupling
with the longitudinal axial vector fluctuations
\bea			\la{act:pseudoscalarm}
 S_{s} &=& - \frac{1}{4 g^2} \int d^5 x \ \sqrt{- G}     \
\lb  \sum_{i=1}^3  \lc \frac{}{} G^{zz} G^{\m \n} \ 
\pa_z \pa_{\m} \chi^i \ \pa_z \pa_{\n} \chi^i  
+ 4 g^2  \ph^2  G^{\m \n} \ \pa_{\m} \chi^i \ \pa_{\n} \chi^i  \rp \rp \nn
&& \qquad \qquad \qquad \qquad  \quad  \quad   \quad  \lp + 4 g^2  \ph^2 \ls G^{zz} \ \pa_z \pi^i \ \pa_z \pi^i 
+ G^{\m\n} \ \pa_{\m} \pi^i \ \pa_{\n} \pi^i \rs  \fr{}{} \rc \nn
&& \qquad \qquad \qquad \qquad  \quad+ 4 g^2  \ph^2  \lc \fr{}{} \ G^{00} \ls V_0^3 \rs^2 \  \ls \ls \pi^1 \rs^2 + \ls \pi^2 \rs^2 \rs
+  2 \ G^{00} \ V_0^3 \ls \pi^1  \ \pa_0 \pi^2 - \pi^2  \ \pa_0 \pi^1  \rs \rp \nn
&& \qquad \qquad \qquad \qquad  \quad \qquad  \quad   - \lp \lp  2 \ G^{\m\n} \sum_{i=1}^{3} \pa_{\m} \chi^i \ \pa_{\n} \pi^i
- 2 \ G^{00} \ V_0^3 \  \ls \pi^1 \ \pa_0 \chi^2  - \pi^2 \ \pa_0 \chi^1  \rs \rc \rb.
\eea
Equations of motion for the scalar fields are summarized as
\bea			\la{eq:neutralscalar}
 \chi^0 - \pi^0  &=& \frac{1}{ 4 g^2  \ph^2 \sqrt{-G} G^{00} } \ \pa_z 
\ls \sqrt{-G} G^{zz} G^{00} \pa_z \chi^0 \rs , \nn
\o_0^2 \ls \chi^0  -  \pi^0  \rs &=& - \frac{1}{\ph^2 \sqrt{-G} G^{00} } 
\ \pa_z \ls \ph^2 \sqrt{-G} G^{zz} \pa_z  \pi^0 \rs ,
\eea
and 
\bea                 \la{eq:chargedscalar}
\o_{\pm} \chi^{\pm} - \ls \o_{\pm} \mp V_0^3  \rs \pi^{\pm} 
&=& \frac{\o_{\pm} }{4 g^2  \ph^2 \sqrt{-G} G^{00}} \pa_z \ls \sqrt{-G} G^{zz} G^{00}
\pa_z \chi^{\pm} \rs , \nn 
\ls \o_{\pm} \mp V_0^3 \rs \lb \o_{\pm}  \chi^{\pm} - \ls \o_{\pm} \mp V_0^3  \rs \pi^{\pm} \rb
&=&   - \frac{1}{\ph^2 \sqrt{-G} G^{00}} \pa_z \ls \ph^2 \sqrt{-G} G^{zz} \pa_z
\pi^{\pm} \rs ,
\eea
where we define
\bea
\pi^0 = \pi^3 \quad , \quad \pi^{\pm} =  \frac{1}{\sqrt{2}} \ls \pi^1 \pm i \pi^2 \rs 
\quad , \quad
\chi^0 = \chi^3 \quad {\rm and} \quad \chi^{\pm} =  \frac{1}{\sqrt{2}} 
\ls \chi^1 \pm i \chi^2 \rs ,
\eea
with
\bea
\pi^0 (t,z) &=& \int \frac{d \o_0}{2 \pi} \ e^{- i \o_0 t} \ \pi^0 (\o_0,z) , \nn
\pi^{\pm} (t,z) &=& \int \frac{d \o_{\pm}}{2 \pi} \ e^{- i \o_{\pm} t} \ \pi^{\pm} (\o_{\pm},z) , \nn
\chi^0 (t,z) &=& \int \frac{d \o_0}{2 \pi} \ e^{- i \o_0 t} \ \chi^0 (\o_0,z) , \nn
\chi^{\pm} (t,z) &=& \int \frac{d \o_{\pm}}{2 \pi} \ e^{- i \o_{\pm} t} \ \chi^{\pm} (\o_{\pm},z) .
\eea

In general, the equation for $\chi$ or $\pi$ can be rewritten as a fourth order differential equation. However, since $V^3_0$ is a constant in the isospin matter, the governing equation further reduces to a third order differential equation. The equation of $\pi$ 
in the isospin matter leads to
\cite{Albrecht:2010eg}
\bea			\la{eq:isospinscalar}
0 &=& \pa_z \lb \frac{1}{\ph^2 \sqrt{-G} G^{00}} \pa_z \ls  \ph^2 \sqrt{-G} G^{zz} \pa_z \pi^0 \rs \rb - \ls \o_0^2 + 4 g^2  G_{00} \ph^2 \rs \pa_z \pi^0 , \la{eq:isospinscalarneut} \\ 
0 &=& \pa_z \lb \frac{1}{\ph^2 \sqrt{-G} G^{00}} \pa_z \ls  \ph^2 \sqrt{-G} G^{zz} \pa_z 
\pi^{\pm} \rs \rb - \lb \ls \o_{\pm} \mp V_0^3\rs^2 + 4 g^2  G_{00} \ph^2 \rb \pa_z \pi^{\pm} \la{eq:isospinnscalarch}.
\eea
By comparing these two equations, we can easily see that the relation between masses
of the neutral and charged pions are given by
\bea
m_{\pi^0} &=& \o_0 , \nn
m_{\pi^{\pm}} &=& m_{\pi^0} \pm  \sqrt{2} \ \pi^2   (\m_P - \m_N) ,
\eea
which is also the same form obtained in the cases of vector and axial vector mesons. 
As a result, in the isospin medium the mass difference between a neutral and charged meson
shows a universal feature for all mesons.  

In the nuclear matter, $V_0^0$ and $V_0^3$ are usually given by the functions of 
the radial coordinate. So the equations governing the charged pions 
and the longitudinal mode of 
the axial vector meson \eq{eq:chargedscalar} can not simply reduce to the forms
 \eq{eq:isospinnscalarch} obtained in the isospin matter, whereas
the neutral pion has the same third order differential equation as \eq
{eq:isospinscalarneut} because it does not depend on $V_0^3$.
Here, we try to solve \eq{eq:neutralscalar} and \eq{eq:chargedscalar} numerically 
with the appropriate
boundary conditions. As mentioned before, since they are generally the fourth order differential 
equation, we need four boundary conditions. We first impose two Dirichlet boundary
conditions at the asymptotic boundary, $\chi^{\a} (0) = \pi^{\a} (0) = 0$
where $\a$ means $\pm$ or $0$. Then, the asymptotic forms of solutions have
\bea
\chi^{\a} &=& c^{\a}_{\chi} z^2 + {\cal O} (z^4) , \nn 
\pi^{\a} &=& c^{\a}_{\pi} z^2 + {\cal O} (z^4) ,
\eea
where $c^{\a}_{\chi}$ and $c^{\a}_{\pi}$ are two remaining integration constants.
In order to fix remaining constants, we impose two additional Neumann boundary conditions
at the IR cutoff, $\pa_z \chi^{\a} (z_{IR}) = 0$ and $\pa_z \pi^{\a} (z_{IR})=0$. 
The former is the one widely used in determining the pion spectrum in the zero density limit,
while the latter is new for the fourth order differential equation.
If one numerically solve \eq{eq:neutralscalar} and \eq{eq:chargedscalar} with
above boundary conditions, one can see how the pseudoscalar meson masses depends on
the nucleon density and density difference.
In Fig. 3 we plot the pion masses depending on the total nucleon number density for 
$\a = 1/2$, which shows that the negatively charge pion become very light in the
relatively small nuclear density compared with the vector and axial vector mesons.

\begin{figure}
\vspace{-1cm}
\centerline{\epsfig{file=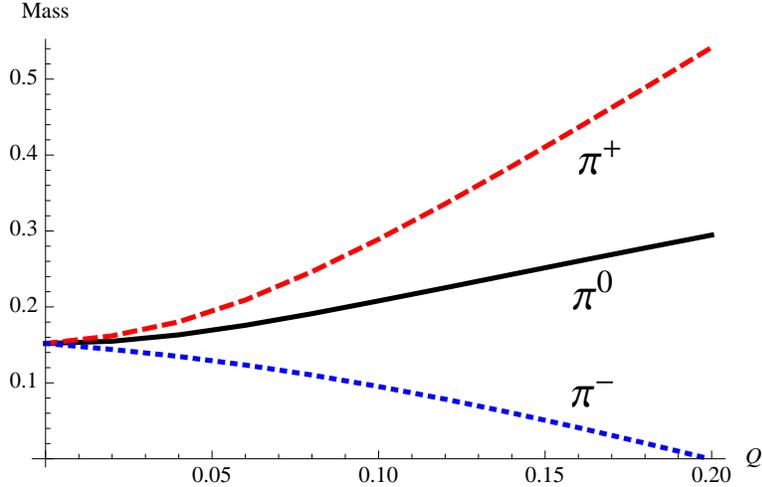,width=10cm}}
\vspace{0cm}
\caption{\small The $a_1$-meson mass splitting for $\a = 1/2$, where we take 
$m_q =2.383 MeV$, $\s = (304 MeV)^3$, $N_c =3$ and $R=1$.}
\label{density}
\end{figure}

Since the deconfinement phase transition occurs at the lower density than the critical point
at which $\pi^-$ becomes massless, there is no pion condensate. However, the critical density
for massless pion is comparable to the critical density of the deconfinement phase transition.
In order to check the pion condensation, we also investigate the $\pi^-$ mass for $\a=1$.
For $\a =1$, the deconfinement phase transition occurs near $Q=0.1615$ while
the massless pion appears near $Q=0.08$ (see Fig. 4). Therefore, pion
can become massless even in the confining phase if $\a$ is relatively large. 
Comparing the data in Fig 3. and 4, we can see that there exists a critical value of $\a$ above which the pion condensation occurs in the nuclear matter.

\begin{figure}
\vspace{-1cm}
\centerline{\epsfig{file=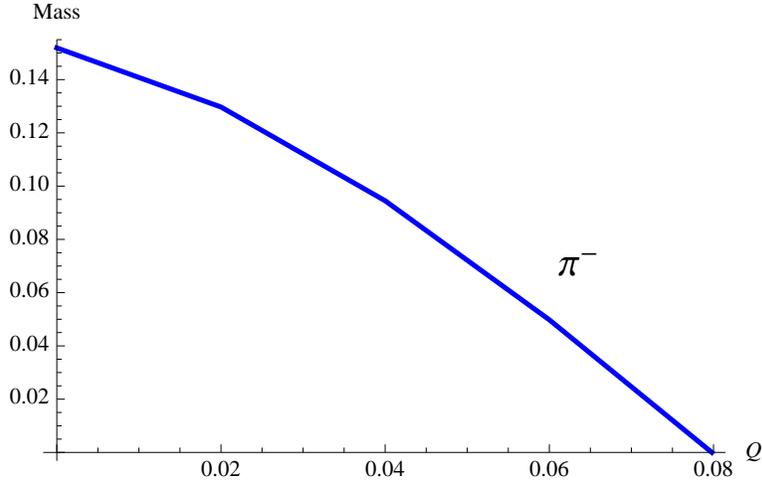,width=10cm}}
\vspace{0cm}
\caption{\small The mass spectrum of $\pi^-$ meson for $a=1$, which shows the
$\pi^-$ becomes massless near $Q=0.08$.}
\label{density}
\end{figure}

\section{The binding energy of a heavy quarkonium in the nuclear matter}

In this section, we will investigate the binding energy of a heavy quarkonium in the
nuclear medium \cite{Gubser:2006bz,Liu:2006nn}. 
The nuclear medium in the confining phase, as shown in the previous section, can be
described by the tcAdS space
\be
ds^2 =  \frac{R^2}{z^2}  \ls - f(z) dt^2 + \frac{1}{f(z)} dz^2  + d \vec{x}^2 \rs ,
\ee
with the metric factor in terms of the nucleon numbers
\bea
f (z) &=& 1 + \D  \ z^6 , \nn
\D &=& \frac{3 \k^2 }{g^2 R^2} \ls Q^2 + \frac{D^2}{9} \rs . 
\eea
On this background, the Nambu-Goto action of an open string 
is given by
\be
S = \frac{1}{2 \pi \a'} \int  d^2 \s \sqrt{ \det
\fr{\pa X^{M}}{\pa \s^{\a}} \fr{\pa X^{N}}{\pa \s^{\b}} G_{MN}} ,
\ee
where $G_{MN}$ is the metric of the tcAdS space. In order to consider a heavy quarkonium
which is a bound state of two heavy quarks, we consider an U-shape open string configuration.
In the static gauge, it can be parameterized by
\bea
\ta &=& t , \quad , \quad \s = x_1=x \quad , \quad x_2 = x_3 = 0 \quad {\rm and } \quad z = z(x) ,
\eea
where $\ta$ and $\s$ are string world sheet coordinates and we assume that the end points of the open string are located at $x = \pm l/2$.
Then, the string action is reduced to
\be
S = \frac{R^2 \b}{2 \pi \a'}   \int_{-l/2}^{l/2} dx \frac{\sqrt{f(z) + z'^2}}{z^2}  ,
\ee
where $\b$ is the appropriate time interval and
the prime means a derivative with respect to $x$. 
By taking the analogy to the particle mechanics
the Hamiltonian, after regarding $x$ as time, is given by
\be \la{ham}
H = - \frac{R^2 \b}{2 \pi \a'} \frac{1}{z^2} \frac{f(z)}{ \sqrt{f(z) + z'^2} } .
\ee
At the tip satisfying $z'=0$, the above conserved Hamiltonian reduces to 
\be \la{hamatmin}
H = - \frac{R^2 \b}{2 \pi \a'} \frac{1}{z_0^2}  \sqrt{f(z_0) }  .
\ee
Comparing these two Hamiltonians, \eq{ham} and \eq{hamatmin},
the inter-quark distance and the binding energy are represented as the functions of $z_0$
\bea
l &=& \- 2 \int_0^{z_0} dz \ \ \frac{\sqrt{f(z_0)}}{\sqrt{f(z)}}
 \frac{z^2 }{\sqrt{f(z) z_0^4 - f(z_0) z^4}} , \nn
V &=& \frac{R^2}{\pi \a'} \int_0^{z_0} dz \ \frac{1}{z^2}
\frac{\sqrt{f(z)}}{\sqrt{f(z)  - f(z_0) z^4/z_0^4}} .
\eea
Since the above binding energy diverges at $z=0$, we need to renormalize it. The 
appropriate counter term is provided by the two straight strings representing the mass 
of two free quarks, whose parameterization is 
\bea
\ta &=& t \quad  , \quad  \s = z \quad  , \quad  x_1 = {\rm const}  \quad {\rm and} \quad x_2 = x_3 = 0 .
\eea
Then, the mass of two free heavy quarks is given by
\be
V_{ct} =  \frac{R^2}{\pi \a'} \int_0^{z_{IR}} dz \ \frac{1}{z^2} ,
\ee
and the resulting renormalized binding energy leads to
\bea \la{binden}
V_{re} &=& V - V_{ct} \nn
&=& \frac{R^2}{\pi \a'} \lb \int_0^{z_0} dz \ \frac{1}{z^2}
\frac{\sqrt{f(z)}}{\sqrt{f(z)  - f(z_0) z^4/z_0^4}}
- \int_{0}^{z_{IR}} dz \ \frac{1}{z^2} \rb .
\eea

In the short inter-distance limit ($ z_0^6 \ll 1/ \D$), the inter-distance and the renormalized
binding energy have the following expansions
\bea
l &=&  A_0 \ z_0  + A_1 \ z_0 \ \D + {\cal O} \ls \D^2 \rs   ,\nn
V_{re} &=& \fr{B_0}{z_0} + B_1  + \fr{B_2 }{z_0} \ \D + {\cal O} \ls \D^2 \rs ,
\eea
where
\bea
A_0 &=& \fr{2 \ \sqrt{\pi} \ \G \ls \fr{7}{4} \rs}{3 \ \G \ls \fr{5}{4} \rs} , \nn
A_1 &=& \fr{25 \ \sqrt{\pi} \ \G \ls \fr{5}{4} \rs}{42 \ \G \ls \fr{3}{4} \rs} 
- \fr{\sqrt{\pi} \ \G \ls \fr{7}{4} \rs}{6 \ \G \ls \fr{5}{4} \rs} , \nn
B_0 &=& - \fr{R^2 \ \G \ls \fr{7}{4} \rs}{3 \ \sqrt{\pi} \ \a' \ \G \ls \fr{5}{4} \rs}, \nn
B_1 &=& \fr{R^2}{\pi \ \a' \ z_{IR}} , \nn
B_2 &=&  \fr{5 \ R^2 \ \G \ls \fr{5}{4} \rs}{12  \ \sqrt{\pi} \ \a'  \ \G \ls \fr{3}{4} \rs} 
- \fr{ R^2 \ \G \ls \fr{7}{4} \rs}{12  \ \sqrt{\pi} \ \a' \ \G \ls \fr{5}{4} \rs} .
\eea
Rewriting the binding energy in terms of the inter-distance leads to
\be
V_{re}  = \fr{A_0 B_0}{l} + B_1 + \fr{A_1 B_0 + A_0 B_2}{l} \ \D + {\cal O} \ls \D^2 \rs .
\ee
Here, $B_1$ is the contribution from the IR cutoff, which is independent of the inter-distance
and the nucleon density. Since the IR cutoff breaks the conformal symmetry,
the binding energy associated with the IR cutoff does not follow the $\fr{1}{l}$ behavior. Ignoring
this part, other terms show the $\fr{1}{l}$-dependence due to the UV conformal symmetry
(see also \cite{Park:2012cu} for the non-conformal and non-relativistic case). 
Since $A_1 B_0 + A_0 B_2$ is always positive, the magnitude of the binding energy decreases
as the total density $Q$ or the density difference $D$ increases. 
This is also true in the long inter-distance region (see Fig. 3 where the binding energy depending on the inter-distance is numerically plotted.). This result implies that
the heavy quarkonium is more easily dissolved in the nuclear medium 
with a large asymmetry of the nucleon number density.

\begin{figure}
\vspace{-1cm}
\centerline{\epsfig{file=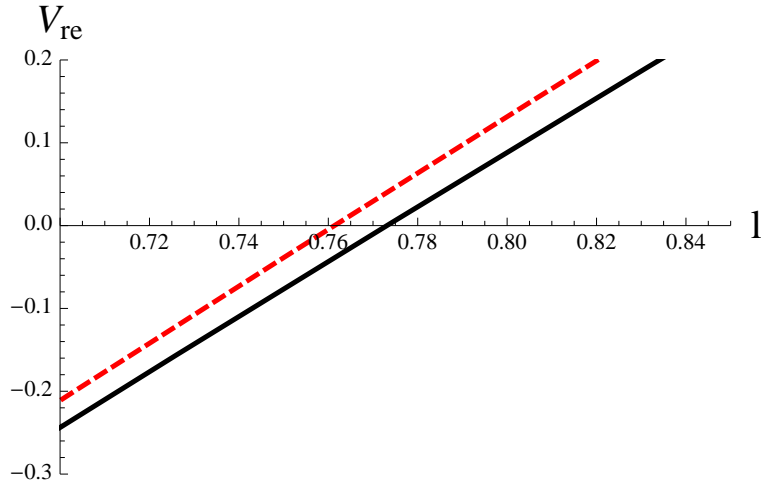,width=10cm}}
\vspace{0cm}
\caption{\small Binding energy of a heavy quarkonium for $\a=0$ (a black solid line) and for $\a=1$ (a red dashed line) where we take $Q=10$, $R=1$ and $\a'=1/\pi$.}
\label{density}
\end{figure}


\section{Discussion}

In this paper, we have studied the meson mass splitting in the nuclear medium
by using the holographic model, the so-called hard wall model. In order
to describe the nuclear matter, we turned on the diagonal components of the
$U(2) \times U(2)$ flavor group, whose combinations describe u- and d-quarks.
In this background, the confinement appears due to the hard wall introduced as an
IR cutoff. After reinterpreting the quark density as the nucleon density in the confining
phase, we investigated the meson masses depending on the total nucleon density and
the density difference. At a given volume, the density difference represents the
asymmetry of the nucleon numbers, which generates the symmetry
energy of the nuclear matter \cite{Park:2011zp} and also the meson mass splitting.

The mass splitting of all mesons in the isospin matter depends linearly on the difference of the isospin chemical potentials. 
In the nuclear matter with a given $\a$ $(= D/Q)$, the masses of the 
negatively charged mesons decrease with the increasing total nucleon density, while 
the masses of the positively charged mesons increase more rapidly.
The numerical study of the meson mass splitting shows that there exists a critical
density in which the negatively charged meson mass becomes zero if there is no deconfinement phase transition. Following \cite{Lee:2009bya,Park:2009nb,Park:2011zp}, 
we showed with $\a =1/2$ that the deconfinement phase transition at zero temperature
occurs before mesons become massless. On the other hand, pion above
a critical value of $\a$ is condensed even in the nuclear matter.

Finally, we also investigated the binding energy of the heavy quarkonium in the nuclear matter.
The asymmetry of the nucleon number density make the binding energy of the quarkonium
slightly weak.

\vspace{1cm}

{\bf Acknowledgements}

C.P. is grateful to S. Shin for her early collaboration and
thanks K. Ulker for his hospitality during the visit to Mimar Sinan Fine Arts University.
This work was supported by the National Research Foundation of Korea(NRF) grant funded by
the Korea government(MEST) through the Center for Quantum Spacetime(CQUeST) of Sogang
University with grant number 2005-0049409. C.P. was also
supported by Basic Science Research Program through the
National Research Foundation of Korea(NRF) funded by the Ministry of
Education, Science and Technology(2010-0022369). Sh.M. was supported by the grant 2221 of TUBITAK of Turkey.

\vspace{1cm}



\begin{thebibliography}{99}

\bibitem{Maldacena:1997re}
  J.~M.~Maldacena,
  Adv.\ Theor.\ Math.\ Phys.\  {\bf 2}, 231 (1998)
  [Int.\ J.\ Theor.\ Phys.\  {\bf 38}, 1113 (1999)]
  [arXiv:hep-th/9711200].

\bibitem{Gubser:1998bc} 
  S.~S.~Gubser, I.~R.~Klebanov and A.~M.~Polyakov,
  Phys.\ Lett.\ B {\bf 428}, 105 (1998)
  [hep-th/9802109].
  
\bibitem{Witten:1998qj} 
  E.~Witten,
  Adv.\ Theor.\ Math.\ Phys.\  {\bf 2}, 253 (1998)
  [hep-th/9802150].

\bibitem{Aharony:1999ti} 
  O.~Aharony, S.~S.~Gubser, J.~M.~Maldacena, H.~Ooguri and Y.~Oz,
  Phys.\ Rept.\  {\bf 323}, 183 (2000)
  [hep-th/9905111].

\bibitem{Klebanov:2000me} 
  I.~R.~Klebanov,
  hep-th/0009139.

\bibitem{Horowitz:2006ct} 
  G.~T.~Horowitz and J.~Polchinski,
  In *Oriti, D. (ed.): Approaches to quantum gravity* 169-186
  [gr-qc/0602037].

\bibitem{Natsuume:2007qq} 
  M.~Natsuume,
  hep-ph/0701201.

\bibitem{Erdmenger:2007cm} 
  J.~Erdmenger, N.~Evans, I.~Kirsch and E.~Threlfall,
  Eur.\ Phys.\ J.\ A {\bf 35}, 81 (2008)
  [arXiv:0711.4467 [hep-th]].

\bibitem{Sachdev:2010ch} 
  S.~Sachdev,
  Lect.\ Notes Phys.\  {\bf 828}, 273 (2011)
  [arXiv:1002.2947 [hep-th]].
  
\bibitem{Erlich:2005qh} 
  J.~Erlich, E.~Katz, D.~T.~Son and M.~A.~Stephanov,
  Phys.\ Rev.\ Lett.\  {\bf 95}, 261602 (2005)
  [hep-ph/0501128].
  
\bibitem{Karch:2006pv} 
  A.~Karch, E.~Katz, D.~T.~Son and M.~A.~Stephanov,
  Phys.\ Rev.\ D {\bf 74}, 015005 (2006)
  [hep-ph/0602229].
  
\bibitem{Sakai:2004cn} 
  T.~Sakai and S.~Sugimoto,
  Prog.\ Theor.\ Phys.\  {\bf 113}, 843 (2005)
  [hep-th/0412141].
  
\bibitem{Sakai:2005yt} 
  T.~Sakai and S.~Sugimoto,
  Prog.\ Theor.\ Phys.\  {\bf 114}, 1083 (2005)
  [hep-th/0507073].
  
\bibitem{Da Rold:2005zs} 
  L.~Da Rold and A.~Pomarol,
  Nucl.\ Phys.\ B {\bf 721}, 79 (2005)
  [hep-ph/0501218].
  
\bibitem{DaRold:2005vr} 
  L.~Da Rold and A.~Pomarol,
  JHEP {\bf 0601}, 157 (2006)
  [hep-ph/0510268].
  
\bibitem{Ghoroku:2005kg} 
  K.~Ghoroku and M.~Yahiro,
  Phys.\ Rev.\ D {\bf 73}, 125010 (2006)
  [hep-ph/0512289].
  
\bibitem{Herzog:2006ra} 
  C.~P.~Herzog,
  Phys.\ Rev.\ Lett.\  {\bf 98}, 091601 (2007)
  [hep-th/0608151].
  
\bibitem{Domokos:2007kt} 
  S.~K.~Domokos and J.~A.~Harvey,
  Phys.\ Rev.\ Lett.\  {\bf 99}, 141602 (2007)
  [arXiv:0704.1604 [hep-ph]].
  
\bibitem{Gursoy:2007cb} 
  U.~Gursoy and E.~Kiritsis,
  JHEP {\bf 0802}, 032 (2008)
  [arXiv:0707.1324 [hep-th]];
  U.~Gursoy, E.~Kiritsis and F.~Nitti,
  JHEP {\bf 0802}, 019 (2008)
  [arXiv:0707.1349 [hep-th]].

\bibitem{Kim:2007em} 
  Y.~Kim, B.~-H.~Lee, S.~Nam, C.~Park and S.~-J.~Sin,
  Phys.\ Rev.\ D {\bf 76}, 086003 (2007)
  [arXiv:0706.2525 [hep-ph]].

\bibitem{Lee:2009bya} 
  B.~-H.~Lee, C.~Park and S.~-J.~Sin,
  JHEP {\bf 0907}, 087 (2009)
  [arXiv:0905.2800 [hep-th]].
  
\bibitem{Park:2009nb} 
  C.~Park,
  Phys.\ Rev.\ D {\bf 81}, 045009 (2010)
  [arXiv:0907.0064 [hep-ph]].
  
\bibitem{Fadafan:2012qy} 
  K.~B.~Fadafan, E.~Azimfard and ,
  Nucl.\ Phys.\ B {\bf 863}, 347 (2012)
  [arXiv:1203.3942 [hep-th]].
  
\bibitem{Jo:2009xr} 
  K.~Jo, B.~-H.~Lee, C.~Park and S.~-J.~Sin,
  JHEP {\bf 1006}, 022 (2010)
  [arXiv:0909.3914 [hep-ph]].

\bibitem{Park:2011zp} 
  C.~Park,
  Phys.\ Lett.\ B {\bf 708}, 324 (2012)
  [arXiv:1112.0386 [hep-th]].
  
\bibitem{Albrecht:2010eg}
  D.~Albrecht and J.~Erlich,
  Phys.\ Rev.\  D {\bf 82}, 095002 (2010)
  [arXiv:1007.3431 [hep-ph]].
  
\bibitem{Son:2000xc} 
  D.~T.~Son and M.~A.~Stephanov,
  Phys.\ Rev.\ Lett.\  {\bf 86}, 592 (2001)
  [hep-ph/0005225].
 
\bibitem{Colangelo:2010pe} 
  P.~Colangelo, F.~Giannuzzi and S.~Nicotri,
  Phys.\ Rev.\ D {\bf 83}, 035015 (2011)
  [arXiv:1008.3116 [hep-ph]];
  C.~Park, D.~-Y.~Gwak, B.~-H.~Lee, Y.~Ko and S.~Shin,
  Phys.\ Rev.\ D {\bf 84}, 046007 (2011)
  [arXiv:1104.4182 [hep-th]];
  P.~Colangelo, F.~Giannuzzi, S.~Nicotri and V.~Tangorra,
  Eur.\ Phys.\ J.\ C {\bf 72}, 2096 (2012)
  [arXiv:1112.4402 [hep-ph]];
  P.~Colangelo, F.~Giannuzzi and S.~Nicotri,
  JHEP {\bf 1205}, 076 (2012)
  [arXiv:1201.1564 [hep-ph]].
  
\bibitem{Parnachev:2007bc} 
  A.~Parnachev,
  JHEP {\bf 0802}, 062 (2008)
  [arXiv:0708.3170 [hep-th]].
  
\bibitem{Kim:2007gq} 
  K.~-I.~Kim, Y.~Kim and S.~H.~Lee,
  arXiv:0709.1772 [hep-ph].

  
\bibitem{Aharony:2007uu} 
  O.~Aharony, K.~Peeters, J.~Sonnenschein and M.~Zamaklar,
  JHEP {\bf 0802}, 071 (2008)
  [arXiv:0709.3948 [hep-th]].

  
\bibitem{Lee:2010dh} 
  B.~-H.~Lee, C.~Park and S.~Shin,
  JHEP {\bf 1012}, 071 (2010)
  [arXiv:1010.1109 [hep-th]];
  C.~Park, B.~-H.~Lee and S.~Shin,
  Phys.\ Rev.\ D {\bf 85}, 106005 (2012)
  [arXiv:1112.2177 [hep-th]].
  
\bibitem{Gubser:2006bz} 
  S.~S.~Gubser,
  Phys.\ Rev.\ D {\bf 74}, 126005 (2006)
  [hep-th/0605182].

\bibitem{Liu:2006nn} 
  H.~Liu, K.~Rajagopal and U.~A.~Wiedemann,
  Phys.\ Rev.\ Lett.\  {\bf 98}, 182301 (2007)
  [hep-ph/0607062].
  
\bibitem{Park:2012cu} 
  C.~Park,
  arXiv:1209.0842 [hep-th];
  C.~Park, 
  arXiv:1305.6690 [hep-th].


  

  
\end{thebibliography}
\end{document}